\documentstyle[11pt,aaspp4]{article}
\newcommand\beq{\begin{equation}}
\newcommand\eeq{\end{equation}}

\begin{document}

\title{Testing the multiple supernovae versus $\gamma$-ray burst
scenarios for giant HI supershells}
\author{Rosalba Perna\altaffilmark{1} and John Raymond}
\medskip
\affil{Harvard-Smithsonian Center for Astrophysics, 60 Garden Street,
Cambridge, MA 02138}
\altaffiltext{1}{Harvard Junior Fellow}
\begin{abstract}

The energy source of the giant HI supershells in nearby galaxies and
in our own is still an unsettled issue. Proposed scenarios for
production of near-complete ringlike supershells are multiple
supernovae (SNe) and $\gamma$-ray bursts (GRBs).  In the late phase
of evolution it is difficult to tell these models apart. We show that,
if a supershell has been powered by multiple SNe, peculiar metal
abundances should be found in the medium within the bubble.  We
present line diagnostics that could detect such unusual abundances,
such as unusually high ratios of [O]/[Fe] and [Ne]/[Fe].
Among ions of the same element, a higher enhancement is
expected in lines from a high-ionization state than
in lines from a low-ionization state. 
Identification of the energy
source of HI supershells would set strong constraints on the rates and
energetics of GRBs, as well as on their location within a galaxy.

\end{abstract}


\section{Introduction}

For several decades, 21 cm surveys of spiral galaxies have revealed the
puzzling existence of expanding giant HI supershells (see e.g. Tenorio-Tagle
\& Bodenheimer 1988 for a review). These nearly spherical structures have 
very low density  in their interiors and high HI
density at their boundaries, and they expand at velocities of several tens of
${\rm km~{s}^{-1}}$.  The radii of these shells are much larger than those
of ordinary supernova remnants and often  exceed $\sim 1$ kpc;  
their ages are typically in the range of $10^6$--$10^8$ years. 
Heiles (1979) denominated as supershells the ones whose
inferred kinetic energies are $\ga 3\times 10^{52}$ ergs.  
The Milky Way  contains  several tens
of them (Heiles 1979; Heiles, Reach, \& Koo 1996), and in one case
the estimated kinetic energy is as high as $\sim 10^{54}$ ergs.  Similar
supershells are also observed in other nearby galaxies.

Whereas it is clear that these HI supershells result from deposition
of an enormous amount of energy in the interstellar medium, 
the energy source is still a subject of
debate. Collisions with high-velocity clouds (Tenorio-Tagle 1981)
could account for those cases where only one hemisphere is present, and
the required input energy is not too large.  However, it is unclear
how such collisions could produce the near-complete ringlike
appearance observed in some cases (Rand \& van der Hulst 1993).

Small shells of radii $\sim 200$--400 pc and energies $\la 3\times
10^{52}$ ergs are often explained as a consequence of the collective
action of stellar winds and supernova explosions originating from OB
star associations (McCray \& Kafatos 1987; Shull \& Saken 1995).  The
winds from the stars of the association create a bubble in the
interstellar medium (ISM) that is filled with hot gas. The bubble
further grows when the stars explode as supernovae, releasing their
energy into the ISM.  Multiple SN explosions are in principle a viable
scenario even for the largest supershells, although this would require very
large OB associations, not typically observed in nearby galaxies
(Kennicutt, Edgar \& Hodge 1989).

Another possibility that has been put forward is that giant supershells
could be the remnants of gamma-ray bursts (GRBs) (Efremov,  Elmegreen \& Hodge 1998;
Loeb \& Perna 1998). In fact, if GRBs occur in galaxies and can have energies
$\ga 10^{53}$ ergs, then remnants in the form of giant bubbles are unavoidable. 
Notice, however, that this conclusion relies on the assumption that the ratio
of $\gamma$-ray energy to kinetic energy of the ejecta is very small, as
required by the popular 'internal shock' models for GRBs. If, on the
other hand, this were not the case, as the analysis of GRB 970508 seems to imply,
then the kinetic energy of GRBs would not be sufficient to produce a giant
remnant (Paczy\'nski 1999).

The nature of the energy source can be more easily identified in
young supershells.  The ones due to multiple SNe would still show
ongoing activity. Bubbles powered by a GRB explosion could instead be
identified by signatures of the radiation emitted by the cooling gas,
which had been heated and ionized by the GRB afterglow (Perna, Raymond
\& Loeb 2000). However, after a time $t\ga 10^5$ yr, the imprints of
this radiation have faded away. Old supershells remain, therefore, the
most difficult to understand\footnote{Among the observed supershells only
about 10\% of them seem to contain OB associations and could therefore
be more naturally attributed to multiple SNe.}. However, given their
ages, they are by far the most abundant in galaxies.  An attempt to
identify their energy source has been recently made by Rhode et
al. (1999). Assuming that the HI holes are created by multiple SNe,
and that the SNe represent the high-mass population (OB stars) of a
cluster with a normal initial mass function, they observed that the
upper main-sequence stars (late B, A and F) should still be present in the
cluster. However, their observations showed that in several of the holes
the observed upper limits for the remnant cluster brightness are inconsistent 
with the expected values. Therefore their test suggested problems with the
multiple SNe scenario. On the other hand, no evidence that the holes could be
due to GRBs was found either. More recently, Efremov, Ehlerova \& Palous (1999)
discussed possible differences between the structures produced by a GRB and by an OB
association, based on their shapes, expansion velocities, and fragmentation times.

Here we propose a new way of testing the multiple SNe versus GRB model to
power supershells. This is based on the fact that SNe inject
metals in the ISM in which they explode. As a result, if a supershell
has been powered by multiple SNe, the abundances of some specific
metals in its interior should be enhanced with respect to the typical
values in the ISM surrounding the shell\footnote{This is commonly
observed in young supernova remnants (e.g. Canizares \& Winkler 1981).}.
As the high-mass stars which power the supershell
explode as Type II SNe, the enhancement should be particularly
pronounced in elements such as Oxygen, Silicon, Neon, Magnesium, but
not in others (e.g. Nomoto et al. 1997).  We present line diagnostics
that could help detect such unusual abundances. 

If a supershell has been powered by a GRB, on the other hand, no
peculiar metal enhancement is expected. The highly relativistic
expansion of the ejecta requires that the baryonic load be very small
\footnote{Even if GRBs were associated 
with SNe (as it has been suggested in the case of SN 1998bw [Galama et al. 1998]),
and there were some mass ejected at later times, it would be just
that of a single SN, and therefore it would be highly diluted within
the large volume of the supershell.}($M\la 10^{-4}M_\odot$). Therefore,
detection of peculiar abundances in the medium within a supershell could 
provide a clue to was the energy source that powered it.
Knowledge of  the fraction of HI supershells  that is likely
to be associated to a GRB event would lead to important constraints on the
energetics and rates of GRBs, as well as on their location within a galaxy.

\section{Evolution of the remnant and metal injection}

We consider a model in which the ambient ISM consists of gas of
uniform density $n_0$, and treat the dynamical evolution of the
supershell in a similar fashion to that of supernova remnants (SNRs).
Whereas the initial stages are very complex and 
depend on details of environment and on how the energy is injected, the
late phases of the evolution are very similar in the two cases
and don't depend much on details. The late evolutionary phases are
what we are interested in. 

Once the mass of the swept-up material exceeds the initial mass of the ejecta
(but while radiative losses are still negligible by comparison with the
initial energy),
the remnant enters a phase of adiabatic expansion (Spitzer
1978). This is described by the self-similar solution derived by
Sedov:
\beq
R_s=1.15\left[\frac{E_0t^2}{\rho_0}\right]^{1/5}\;,
\label{eq:rs}
\eeq
where $R_s$ is the radius of the shock, $E_0$ the energy of the explosion,
and $\rho_0=\mu m_p n_0$. Let $r$ be the radial coordinate
in the interior of the remnant, $x\equiv r/R_s$, and $v_s$ the shock velocity;
then the density profile can be approximated analytically  by (Cox \& Anderson 1982)
\beq
\rho=4\rho_0\left[\frac{5}{8}+\frac{3}{8}x^8\right]x^{4.5}\exp\left[-\frac{9}{16}
(1-x^8)\right]\;,
\label{eq:n} 
\eeq
the pressure by
\beq
p=\frac{3}{4}\rho_0v_s^2\left[\frac{5}{8}+\frac{3}{8}x^8\right]^{5/3}
\exp\left[-\frac{3}{8}(1-x^8)\right]\;,
\label{eq:p}
\eeq
and the velocity of an element of the shell at position $x$ by 
\beq
v=\frac{3}{4}v_s\frac{x}{2}\left[\frac{1+x^8}{\frac{5}{8}+\frac{3}{8}x^8}\right]\;. 
\label{eq:v}
\eeq
The temperature in the interior is then found from Equations (\ref{eq:n})
and (\ref{eq:p}) and the use of the equation of state, $p=nkT$.

The remnant continues to expand adiabatically up to the time at which
radiative cooling begins to dominate, that is when the gas temperature
behind the front reaches the value $T\approx(5-6)\times 10^5$~K,
corresponding to the maximum of the cooling curve.  By the time the
remnant has arrived at this stage, approximately half of its thermal
energy has been radiated away, and a cold dense shell is formed,
containing about half of the mass of the swept-up gas. The cavity
bounded by this shell contains hot, low-density gas that continues to
expand nearly adiabatically (Lozinskaya 1992; Cui \& Cox 1992 for the
cases where thermal conduction is neglected).  The evolution of the
remnant following the formation of the cold shell is well described 
by a pressure-driven snowplough (PDS; Cox 1972; McKee \& Ostriker
1977; Lozinskaya 1992).  The time at which the PDS phase starts is
given by Cioffi, McKee \& Bertschinger (1988), 
\beq t_{\rm
PDS}=4\times 10^4 E^{0.23}n_0^{-0.3}\eta^{-0.35}\,{\rm yr}\;,
\label{eq:tPDS}
\eeq 
where $\eta$ is the metallicity ($\eta=1$ for solar abundances).
This value is similar to that derived by other authors (e.g. Chevalier
1974; Falle 1981). Differences are mainly due to the use of different
cooling functions, although the shell velocity predicted at the same
radius is very similar for all calculations.  In the PDS phase,
the radius of the shell evolves as $R_s\propto t^{0.31}$ (Chevalier
1974).    During this  phase
the remnant radiates most of its energy away, and therefore the physical
variables describing the shell evolution are expected to
deviate from the self-similar solution given in Equations
(\ref{eq:n}--\ref{eq:v}) (see eg. Weaver, McCray \& Castor 1977).  However, as
already mentioned, when 
thermal conduction is neglected the hot center of the bubble continues to
evolve nearly adiabatically, though cooling eats into it from the
outside.  Here we only consider lines arising from gas at temperatures
$\ga 10^6$ K, for which cooling is not yet important, and therefore
it is a good approximation to  
keep the evolution as given by Equations (\ref{eq:n}--\ref{eq:v}) in the
interior of the bubble,  where these lines are produced. 
Moreover, notice that metallicity effects  can significantly alter 
the early evolution of the remnants; however during the late PDS phase the differences
due to metallicity are found to be negligible (Thorton et al. 1998; Goodwin, Pearce
\& Thomas 2000). 

Observations of supershells yield their radii and expansion
velocities, from which their kinetic energies can be inferred
(e.g. Heiles 1979). The kinetic energy, however, is only a small
fraction of the total energy released in the ISM. A large fraction of
the energy is, in fact, radiated away by the cooling bubble.
Numerical simulations of supernova explosions show that only a
fraction $f\la 4\%$ of the energy of the explosion is found as
kinetic energy in the very late phase of evolution of the
remnant (Chevalier 1974; Goodwin, Pearce \& Thomas 2000). Therefore,
if an old supershell has an inferred kinetic energy $E_{\rm K}$, 
and its energy input  is provided by multiple SNe, the number of SNe required is
$N_*\approx E_{\rm K}/(fE_{\rm SN})$. The energy released by a SN
explosion is typically taken to be $E_{\rm SN}=10^{51}$ ergs, in
agreement with the value inferred from the modelling of SN 1987A and SN 1993J
(Shigeyama \& Nomoto 1990; Shigeyama et al. 1994).
However, before
they explode, the most massive stars of the OB association contribute
to the mechanical energy of the bubble with their winds (McCray \&
Kafatos 1987; Heiles 1987; Shull \& Saken 1995). The wind energy
varies with optical luminosity (Abbott 1982), but, as an average, 
Heiles (1987) assumes a value  of $1.17\times 10^{51}$ ergs
per star. This brings the number of required stars to $N_*\approx
E_{\rm K}/(fE_{\rm SN})$ with $f\approx 0.05$.  Because of the
uncertainties in these estimates, we prefer to adopt
a more conservative value, and therefore we take  $f=10\%$ in our
calculations.

If multiple SNe have provided the power for 
a supershell of energy $E_{\rm K}$, we
assume that initially there was a cluster of $N_*$ OB stars
distributed according to an initial mass function (IMF). The IMF for
such stars can be written as (Garmany, Chionti \& Chiosi 1982)
\beq 
f_{\rm IMF}(M_*)\equiv dN_*/dM_*\propto M_*^{-\beta}\;,  
\label{eq:IMF}
\eeq
where $\beta\sim 2.0-2.7$. Here we adopt
$\beta=2.3$, and normalize the distribution so that $\int_{M_{\rm
min}}^{M_{\rm max}} f_{\rm IMF}(M_*)dM_*=N_*$.

The main-sequence lifetimes of massive stars are given
approximately by (Stothers 1972; Chiosi,
Nasi, \& Sreenivasan 1978)
\beq
t_*\sim \left\{
  \begin{array}{ll}
3\times 10^7(M_*/10M_\odot)^{-1.6}\; {\rm yr} \;\;
& \hbox{if $\;7\;\la M_*\la 30 M_\odot$}\\   
 9\times 10^6(M_*/10M_\odot)^{-0.5} \;{\rm yr} \;\;
& \hbox{if $\;\;30\la M_*\la 80 M_\odot$} \\
\end{array}\right.\;.
\label{eq:tMS}
\eeq 

The least massive star that is expected to terminate as a Type II SN
has initial mass $M_{\rm min}=7 M_\odot$ (Trimble 1982). We take
$M_{\rm max}=100 M_\odot$ as the mass of the most massive star of the
association\footnote{We consider a model of an OB 
association with coeval star formation (see e.g. Shull \& Saken 1995);
that is all stars are assumed to be formed at once with no age
spread. We don't expect very sensitive variations in our results with
the introduction of a spread in birth dates, as long as most of the stars
explode in the early phase of the supershell. This assumption is
consistent with the observation that most old supershells do not show any
more signs of an OB association within them.} (Shull \& Saken 1995). 

Metallicity yields in SNe have been obtained in numerical
simulations by a number of authors.
Here we use the results of the computation made by
Nomoto et al. (1997), who  have calculated metallicity yields for
several values of the progenitor mass between 13 and 70 $M_\odot$.  
For other values of masses between our $M_{\rm min}$ and $M_{\rm max}$, 
we interpolate and extrapolate their values.
Metals are injected in the ISM when the stars of the association
become SNe.  Let $X(M_*)$ be the yield of a star of mass $M_*$ in 
element $X$, and let $M_*(t)$ be the mass of a star with lifetime $t$,
as given by  Equation (\ref{eq:tMS}).  
 The amount of mass of element $X$ that is injected in
the ISM between the times $t_1$ and $t_2=t_1+\Delta t$ is then 
\beq
\Delta M_X=\int_{M_*(t_2)}^{M_*(t_1)}f_{\rm IMS}(M_*)X(M_*)dM_*\;. 
\label{eq:delm}
\eeq
We take the initial time of the supershell to be that at which the first
supernova goes off. Our final results are not very sensitive to this
particular choice as long as the time at which the first star explodes
is much smaller than the lifetime of the supershell. We assume
spherical symmetry, and slice the volume of the bubble into a number
of shells.  Each shell is followed in time, and the concentrations of
the ions of each element are computed, allowing for
time-dependent ionization. The stepsize $\Delta t$ is chosen so that $\Delta
t/t\ll 1$ at every time.  After each time increment $\Delta t$, a new
shell is added while the others evolve according to
Equations~(\ref{eq:n}-\ref{eq:v}).  During each $\Delta t$, an amount
of mass $\Delta M_X$ as given by Equation (\ref{eq:delm}) is injected
into the expanding bubble.  How these extra elements precisely mix with the medium in
the supershell is a very complicated problem, and its solution depends on
details of the model.  However, we consider it reasonable to assume that mixing is
negligible between SN material that is injected in the supershell at a
given time, and ISM that is accreted by the supershell at much later
times\footnote{
Notice that the lifetime of each SN (i.e. the time that it takes
the shock to slow down) is much smaller than the
lifetime of the supershell, in the case where $N_*\gg 1$.}.  Within the
supershell, we then make the simplest assumption of uniform mixing of the
ejecta with the medium. 

Finally, a modified version of the Raymond-Smith emission code
(Raymond \& Smith 1977) is used to compute, at each time and for each
shell, ionization and recombination rates, the time-dependent
ionization state, and the $X$-ray spectrum.

\section{Diagnostics of metal enhancements}

Figure 1 shows the enhancements (relative to standard solar values
and for $n_0=1$ cm$^{-3}$)
in the abundances of Oxygen, Silicon
and Neon for a supershell of age $t=5\times 10^7$ yr and kinetic 
energy\footnote{As long as $N_*\gg 1$ (and therefore we can apply our
assumption about mixing) our results at late times are roughly
independent of $E_{\rm K}$ (or equivalently $E_0$). In fact, as $R_s
\propto E_0^{0.32}$, the volume of the shell is $V\propto E_0$. The
mass injected in element $X$ is $\Delta M_X\propto E_0$, and therefore
the number density $n_X=\Delta M_X/V \sim$ independent of $E_0$.} $E_{\rm K} =
5\times 10^{53}$ ergs.  As explained in \S 2, the kinetic energy as
measured at late times is assumed to be on the order of 10\% of the total input energy
in SNe. The total amount of Oxygen mass injected by the SNe is then
$\sim 10^4$ $M_\odot$, for the assumed $E_{\rm K}$.   
The enhancements in the abundances are  more pronounced in the inner regions of
the supershell, as a consequence of the fact that most of the extra
mass is injected at early times, due to the shorter lifetimes of the
most massive stars. Notice that, whereas these results are shown for bubbles accreting from
an ISM with solar metallicity, a stronger enhancement in the abundances could
be observed for bubbles growing in a medium with low metallicity, such as
that of the Large Magellanic Cloud, where the abundance of Oxygen is about
half the solar value (Vancura et al. 1992). On the other hand, if mixing 
is more efficient than assumed, then more Oxygen would be found in the 
outer colder regions of the supershell, therefore reducing its enhancement.
As already emphasized, unusual metal enhancements are only
expected in supershells due to multiple SNe, but not in those powered
by GRBs. 

Figure 2 shows the emitted power in  some of the
strongest $X$-ray lines as a function of the position within the supershell. 
This is shown at various ages of
the supershell. Due to the overall cooling of the shell with time, the
hot region ($T\ga 10^6$ K) in which these lines are produced moves
towards its inner part. As a consequence, the enhancements
inferred from measurements of these lines will increase with
time. This is illustrated by  the solid lines of Figure 3, where
several line ratios are shown at various ages of a supershell powered
by multiple SNe. Here we have plotted ratios between lines of elements 
(such as O, Si, Ne, Mg) which are particularly enhanced by the
SN explosion, and lines of elements (such as Fe, N) which are less affected.
The enhancements are best inferred by using ratios of two lines of similar
energy from different elements. These depend on the relative abundances
of the two elements and on the ionization fractions for each element,
but have no other significant dependence on the electron 
temperature\footnote{Ratios between two lines close in energy are not much affected
by interstellar absorption either.} $T$. 
Thus the abundances can be determined once the ionization fractions
are known. These, in turn, can be found from ratios of lines at approximately
the same energy from different ionization stages of the same element. 
In cases where the continuum is observable, measurements of line strengths
relative to the local continuum might permit the determination of absolute
abundances for an ionic species (Winkler et al. 1981).
 
The dotted lines in Figure 3 show the same line ratios used for the
case of multiple SNe, but for standard solar abundances, as they would
appear if the supershell had been powered by a GRB. We find that
enhancements in some specific line ratios by a factor of a few are
expected in a supershell produced by multiple SNe with respect to a
supershell due to a GRB. However, we need to emphasize that the
precise value of the enhancement in each ion of each element will of
course vary depending on the details of mixing within the shell.
Nonetheless, what we hoped to identify are general features that a
supershell due to multiple SNe is expected to have, as opposed to a
supershell powered by a GRB.  That is, a strong enhancement in the
abundances of some specific elements such as O, Si, Ne, Mg, but not
others. Moreover, among ions of the same element, a higher
enhancement is expected to be seen in lines from a high ionization
state as compared to lines from a low ionization state,
the latter being produced in the outer cold shell, which has
most of the mass accreted from the ISM at later times.

An issue that we have neglected in our model is that of thermal conduction
across the interface between the dense outer shell and hot interior. Fast
electrons from the hot interior can penetrate significant distances in
the cold shell before depositing their energy in collisions with the
gas, thus transferring heat across the contact discontinuity. The
resulting heating raises the pressure of the inner edge of the shell,
which then expands into the hot interior. It has been shown that
tangled magnetic fields are able to partially suppress thermal
conduction, but Slavin \& Cox (1993) showed that even a small amount of conduction
can lead to effective cooling in the end. If thermal conduction
operates, in fact, bubbles and superbubbles would be colder but denser
in their interiors, and therefore their $X$-ray emission would be
suppressed.  The importance of the effect of thermal conduction in
bubbles is still an open issue, and the observational evidence appears
mixed indeed. While some bubbles are fainter in $X$-ray than predicted by the
theory, others are brighter, by up to an order of magnitude (Mac Low
1999).  A detailed modelling of the $X$-ray emission under the various circumstances
is not within the scope of our paper, and therefore we have adopted a simple model.

Within the framework of this model, the brightest $X$-ray lines in the
late phase of a supershell of $E_{\rm K}=5\times 10^{53}$ ergs are 
expected to have luminosities in the range of $10^{31} - 10^{32}$ ergs.   
For supershells at galactic distances of
a few kpcs, these lines are within the detection capability of {\em CHANDRA} or {\rm XMM}.
In cases where the emission lines are too faint to be detected, it 
would be useful to probe supershells in absorption.  In fact, given their
sizes on the sky ($\ga$ a few deg$^2$ [Heiles 1979] for those in our galaxy),
it is likely to find a bright source behind them. Metal enhancements could then
be detected by measuring the equivalent widths of absorption lines in the 
spectrum of the source.  Again, it would be useful to compare strengths of 
absorption lines of the most enriched elements with those of elements which are
not affected by SNe yields, and, among ions of the same element, to compare  
strengths of absorption lines from different ionization states. 
It would be worthed to attempt this test, either in emission or in absorption, 
especially with the most energetic supershells. Several have been observed
which require an input energy $\ga 10^{54}$, both in our Galaxy (Heiles 1979),
and in nearby ones, such as, for example, NGC 4631 (Rand \& Van der Hulst 1993) or
NGC 3556 (Giguere \& Irwin 1996).

\section{Conclusions}
   
The energy source which powers giant HI supershells is still a
subject of debate.  Its identification is particularly difficult in
the late phases of evolution of the remnant.
While hemispherical supershells  could be perhaps 
attributed to collisions with high-velocity clouds, the near-complete 
ringlike ones could be more easily explained by either multiple SNe from
an OB association or by a GRB.

In this paper we have identified signatures that could help discriminate
between the two models.  Namely, we have shown that supershells
powered by multiple SNe are likely to show enhanced abundances of the
metals produced by the SNe themselves, and we have proposed some line diagnostics that
could help reveal these unusual features.

Being able to discriminate between the multiple SNe and the GRB scenario  
for the production of HI supershells would help constrain GRB rates and energetics, as
well as their location within a galaxy.

\begin{figure}[t]
\centerline{\epsfysize=5.7in\epsffile{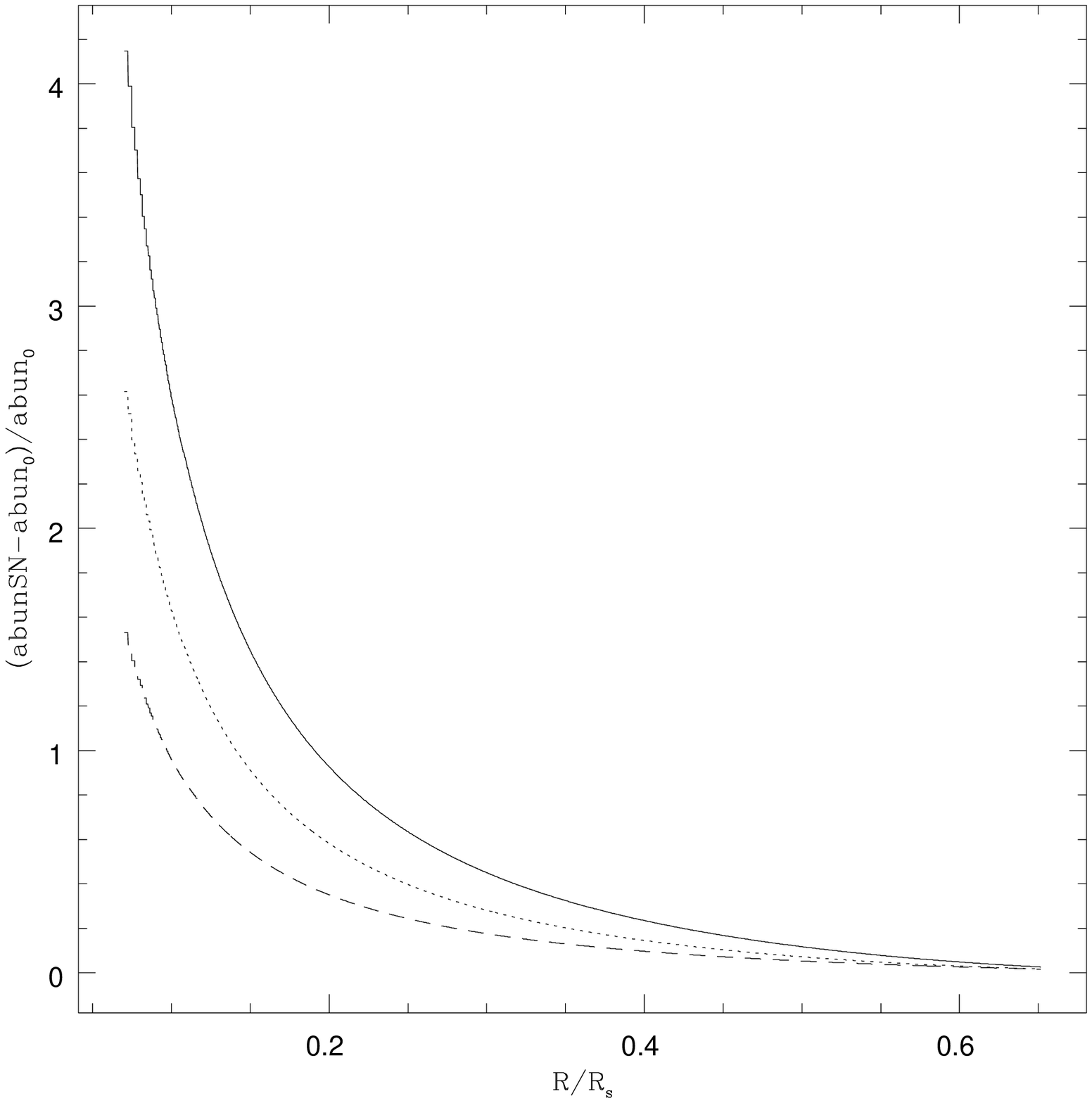}}
\caption{Fractional enhancements in the abundances of Oxygen (solid line), Neon
(dotted line) and Silicon (dashed line) for a supershell which has been
powered by multiple SNe. 
The energy is 
$E_{\rm K}=5\times 10^{53}$ ergs and the age is $t=5\times 10^7$ yr. }
\label{fig:1}
\end{figure}

\begin{figure}[t]
\centerline{\epsfysize=5.7in\epsffile{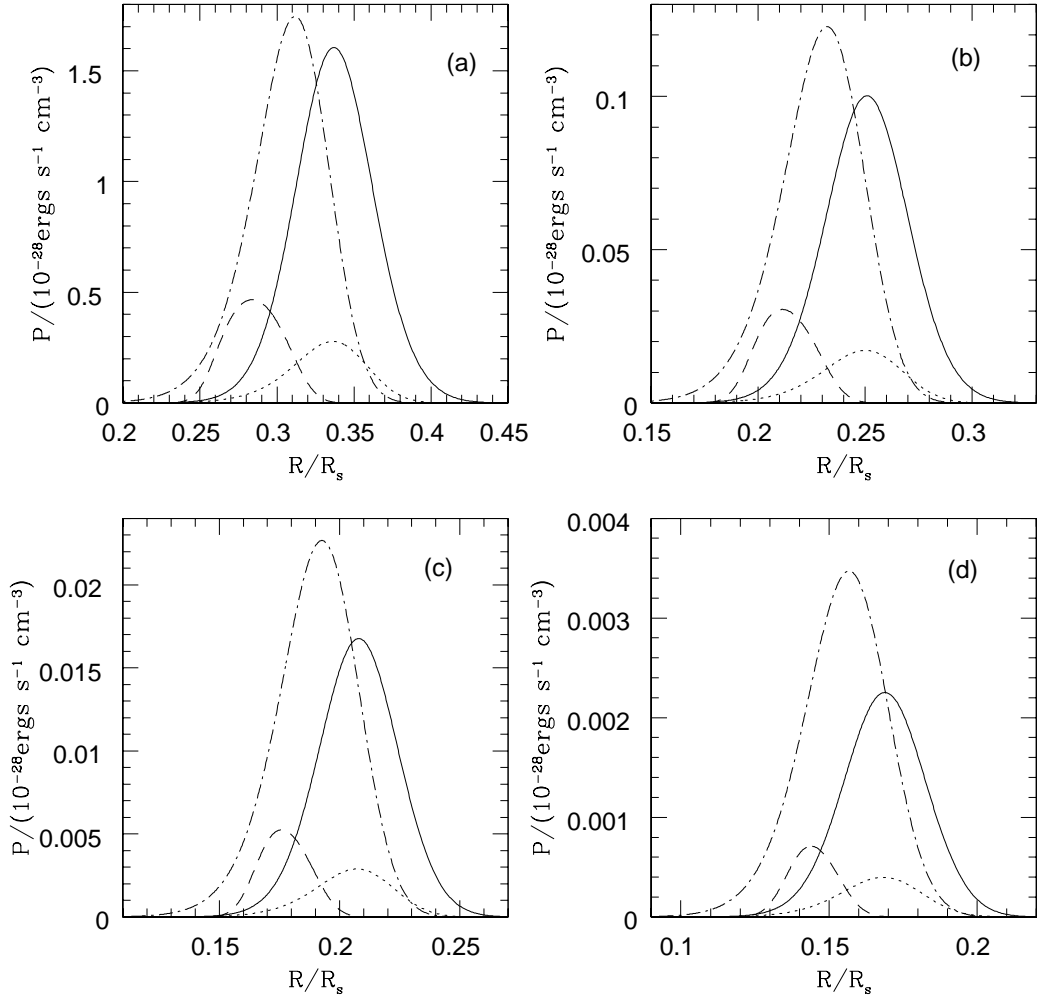}}
\caption{Emitted power at various times in some of the strongest lines:
O VII $\lambda 21.6$ (solid line), Si XIV $\lambda 6.18$ (dotted line),
Fe XVII $\lambda 17.05$ (dashed line), OVIII $\lambda 18.97$ (dotted-dashed line).
The times are $t=3\times 10^6$ yr [panel(a)], $t=10^7$ yr [panel(b)],
$t=3\times 10^7$ yr [panel(c)], $t=10^8$ yr [panel(d)]. }
\label{fig:2}
\end{figure}

\begin{figure}[t]
\centerline{\epsfysize=5.7in\epsffile{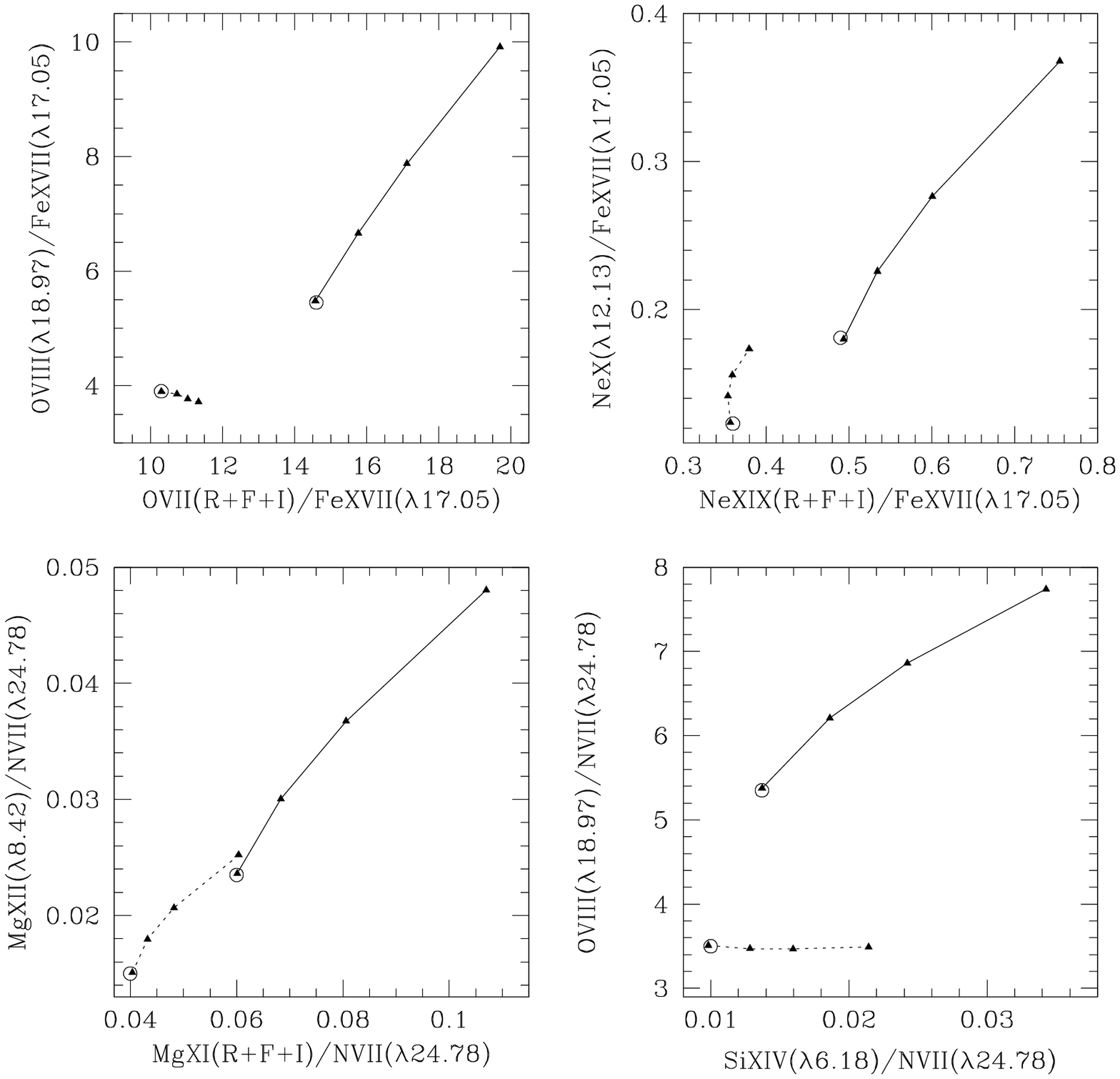}}
\caption{Line ratios that could be diagnostics of unusual metal enhancements
in a supershell powered by multiple SNe (solid lines) with respect to 
a supershell powered by a GRB (dotted lines). The ratios are shown
at the same times as in Figure 2. The earliest time is marked by a circle. 
The symbols R, F and I stand for
resonance, forbidden and intercombination line, respectively. }
\label{fig:3}
\end{figure}


\begin{references}

\reference{}
Abbott, D. C. 1982, ApJ, 263, 723
\reference{}
Canizares, C. R., \& Winker, P. F. 1981, ApJ, 246, L33 
\reference{}
Chevalier, R. A. 1974, ApJ, 188, 501
\reference{}
Chiosi, C., Nasi, E., \& Sreenivasan, S. P. 1978, A\&A, 63, 103
\reference{}
Cioffi, D. F., McKee, C. F.\& Bertschinger, E. 1998, ApJ, 334, 252
\reference{}
Cox, D. P. 1972, ApJ, 178, 159
\reference{}
Cox, D. P. \& Anderson, P. R. 1982, ApJ, 253, 268
\reference{} 
Cui, W. \& Cox, D. P. 1992, ApJ, 401, 206
\reference{}
Efremov, Y. N., Elmegreen, B. \& Hodge, P. W. 1998, ApJ, 501, L163 
\reference{}
Efremov, Y. N., Ehlerova, S., \& Palous, J. 1999, A\&A, 350, 468
\reference{}
Falle, S. A. G. E. 1981, MNRAS, 195, 1011
\reference{}
Galama, T. J. et al. 1998, Nature, 395, 672
\reference{}
Garmany, C. D., Conti, P. S., \& Chiosi, C. 1982, ApJ, 263, 777
\reference{} 
Giguere, D. L. \& Irwin, J. 1996, AAS, 189, 680S
\reference{}
Goodwin, S. P., Pearce, F. R. \& Thomas, P. A., astro-ph/0001180 
\reference{}
Heiles, C. 1979, ApJ, 229, 533
\reference{}
Heiles, C., Reach, W. T, \& Koo, B-C. 1996, ApJ, 466, 191
\reference{}
Heiles, C. 1997, ApJ, 315, 555
\reference{}
Kennicutt, R. C., Edgar, B. K. \& Hodge, P. W. 1989, ApJ, 337, 761
\reference{}
Loeb, A. \& Perna, R. 1998, ApJ, 503, L35
\reference{}
Lozinskaya, T. A. 1992, 'Supernovae and Stellar Wind in the Interstellar Medium',
(AIP: New York)
\reference{} 
Mac Low, M.-M. 1999, astro-ph/9912536
\reference{}
McCray, R. \& Kafatos, M. 1987, ApJ, 317, 190
\reference{}
McKee, C. F. \& Ostriker, J. P. 1977, ApJ, 218, 148
\reference{}
Nomoto, K., Hashimoto, L. Tsujimoto, T., Thielemann, F. K., Kishimoto, N.,
Kubo, Y., \& Nakasato, N. 1997, Nucl. Phys. A616
\reference{}
Paczy\'nski, B. astro-ph/9909048
\reference{}
Perna, R., Raymond, J. \& Loeb, A. 2000, ApJ in press, preprint astro-ph/9904181
\reference{} 
Rand, R. J. \& van der Hulst, J. M. 1993, AJ, 105, 2098
\reference{} 
Raymond, J. \& Smith, B. W. 1977, ApJS, 35, 419
\reference{}
Rhode, K., L., Salzer, J. J., Westpfahl, D. J. \& Radice, L. A. 1999, AJ, 118, 323
\reference{}
Slavin, J. D., Cox, D. P. 1993, ApJ, 417, 187
\reference{}
Spitzer, L. 1978, 'Physical Processes in the Interstellar Medium' (PUP: Princeton)
\reference{}
Shigeyama, T. \& Nomoto, K. 1990, ApJ, 360, 242
\reference{}
Shigeyama, T., Suzuki, T., Kumagai, S., Nomoto, K., Sayo, H., \& Yamaoka, H.
1994, ApJ, 420, 341  
\reference{}
Shull, J. M. \& Saken, J. M. 1995, ApJ, 444, 663 
\reference{}
Stothers, R. 1972, ApJ, 175, 431
\reference{}
Tenorio-Tagle, G. 1981, A\&A, 94, 338
\reference{}
Tenorio-Tagle, G. \& Bodenheimer, P. 1988, ARA\& A, 26, 145
\reference{}
Thorton, K., Gaudlitz, M., Janka, H.-Th. \& Steinmetz, M. 1998, ApJ, 500, 95 
\reference{}
Trimble, V. 1982, Rev. Mod. Phys. 54, 1183 
\reference{}
Vancura, O., Blair, W. P., Long, K, S., \& Raymond, J. C. 1992, ApJ, 394, 158
\reference{}
Weaver, R., McCray, R., \& Castor, J. 1977, ApJ, 218, 377
\reference{}
Winkler, P. F., Canizares, C. R., Clark, G. W., Markert, T. H., \& Petre, R.
1981, ApJ, 245, 574 

\end{references}
\end{document}